\documentclass[
aps,
prc,
superscriptaddress,
floatfix,
nofootinbib,
twocolumn
]{revtex4-2}

\usepackage{amsmath,amssymb,amsfonts,physics,graphicx,float}

\usepackage[setpagesize=false]{hyperref}
\allowdisplaybreaks[4]

\usepackage[labelformat=simple]{subcaption} 

\usepackage{ulem}
\usepackage{color}
\definecolor{ar}{rgb}{1.0, 0.01, 0.24}
\definecolor{al}{rgb}{0.82, 0.1, 0.26}
\definecolor{ev}{rgb}{0.56, 0.0, 1.0}

\begin{document}

\title{
Ferromagnetic instabilities in quarkyonic matter
}

\author{Bikai Gao}
\email{bikai@rcnp.osaka-u.ac.jp}
\affiliation{Research Center for Nuclear Physics, The University of Osaka, Ibaraki, Osaka 567-0047, Japan}

\author{Kenichi Yoshida}
\email{kyoshida@rcnp.osaka-u.ac.jp}
\affiliation{Research Center for Nuclear Physics, The University of Osaka, Ibaraki, Osaka 567-0047, Japan}
\affiliation{RIKEN Nishina Center for Accelerator-Based Science, Wako, Saitama 351-0198, Japan}
\affiliation{Center for Computational Sciences, University of Tsukuba, Tsukuba, Ibaraki 305-8577, Japan}

\date{\today}

\begin{abstract}
We investigate the magnetic properties of quarkyonic matter, which naturally bridges nuclear and quark matter at intermediate densities relevant to neutron star cores. We extend the quarkyonic model to include spin polarization, where nucleons near the Fermi surface can be polarized while quarks in the deep Fermi sea remain unpolarized due to strong Pauli blocking. After including neutron interactions with spin-dependent terms, we find that quarkyonic matter can develop ferromagnetic instabilities at low densities, characterized by negative magnetic susceptibility. This ferromagnetic behavior occurs in pure neutron matter, independent of proton contributions, and results from the competition between attractive spin-dependent interactions and kinetic energy costs. The system returns to paramagnetic behavior at higher densities when Pauli pressure dominates. Our results demonstrate that the splitting of Fermi momenta of quarkyonic matter produces fundamentally different magnetic responses compared to conventional nuclear matter, with important implications for neutron star magnetism and magnetar physics.
\end{abstract}

\maketitle


\section{Introduction}

Neutron stars (NSs) represent one of the most extreme environments in the universe, with central densities that can exceed several times of the nuclear saturation density ($n_0 \approx 0.16$ fm$^{-3}$). Understanding the properties of matter under such extreme conditions remains one of the most fundamental challenges in nuclear physics and astrophysics\cite{Lattimer:2012nd,Hell:2014xva,Baym:2017whm}.

To study the NS, we have to consider its equation of state (EOS) which  can directly determines the mass-radius relationship of NSs, and various observable phenomena \cite{Lattimer:2015nhk, Ozel:2016oaf,Baym:2017whm,Clevinger:2025acg}. However, there are still many difficulties to obtain the EOS across the whole density region relevant to NS. At low densities near nuclear saturation density, matter is well-described by nucleonic degrees of freedom with established nuclear interactions.  At higher densities ($n_B \geq 5n_0$), quarks may become deconfined, requiring consideration of quark degrees of freedom. However, the intermediate density region (from $2n_0$ to $5n_0$) is particularly problematic, as nuclear interactions become increasingly complex while quark degrees of freedom may gradually emerge. Traditional approaches bridge this intermediate gap through extrapolation or interpolation methods, assuming specific scenarios of the quark-hadron transition, such as a smooth crossover \cite{Baym:2017whm, Baym:2019iky,Kojo:2021wax, Minamikawa:2020jfj,Gao:2022klm,Minamikawa:2023eky,Kong:2023nue,Gao:2024chh,Kong:2025dwl} or a first-order phase transition \cite{Benic:2014jia,Christian:2023hez,Gao:2024lzu,Kourmpetis:2025zjz,Verma:2025vkk,Yuan:2025dft,Christian:2025dhe}. However, these assumptions introduce model dependencies that are difficult to quantify.

The quarkyonic matter description \cite{McLerran:2007qj,McLerran:2008ua,Hidaka:2008yy,Fukushima:2015bda,Duarte:2021tsx,Kojo:2021ugu,McLerran:2018hbz,Jeong:2019lhv,Sen:2020peq,Fujimoto:2023mzy,Zhao:2020dvu,Ivanytskyi:2025cnn} offers an
intriguing picture of the transition from hadronic to quark matter. The basic concept of quarkyonic matter is that at sufficiently high baryon chemical potential, the degrees of freedom inside the Fermi sea can be treated as quarks, while confining forces remain important only near the Fermi surface. Nucleons emerge through correlations between quarks at the surface of the quark Fermi sea at high densities \cite{Fujimoto:2023mzy,Zhao:2020dvu,Gao:2024jlp,Fujimoto:2024doc,Ivanytskyi:2025cnn}. Recent observations of NSs have placed stringent constraints on the EOS \cite{Fonseca:2016tux,LIGOScientific:2017vwq,LIGOScientific:2017ync,Miller:2021qha,Riley:2021pdl,Fonseca:2021wxt,Vinciguerra:2023qxq,Kacanja:2024hme}, requiring it to be sufficiently stiff in the high density region to support the massive NS with mass exceeding $2M_{\odot}$, while remaining soft in the hadronic region to produce the stellar radii. The quarkyonic framework has shown promising in satisfying these observational constraints while maintaining consistency with nuclear physics at low densities.

Another complexity to construct the NS EOS arises from the presence of strong magnetic fields in NSs, particularly in magnetars where surface fields can reach $10^{14}$-$10^{15}$ G \cite{Janka:2012wk,Vigano:2013lea,Chatterjee:2014qsa,Dexheimer:2016yqu,Sedrakian:2018ydt,Blaschke:2018mqw}. Such strong magnetic fields can induce spin polarization in dense matter, modifying the EOS and potentially leading to novel phases \cite{Marcos:1991qd,Cardall:2000bs,Maruyama:2000cw,Bordbar:2011rt,Kawaguchi:2018xug,Pons:2019zyc,Kawaguchi:2024edu}. The response of matter to these fields depends critically on the underlying degrees of freedom and their interactions, making the choice of matter description crucial for understanding NS magnetism\cite{Maruyama:2000cw,Tatsumi:2003bk}.

The quarkyonic framework provides a unique perspective for studying magnetized dense matter. Unlike previous investigations that have focused separately on either purely nuclear matter or fully deconfined quark matter, the quarkyonic approach naturally unifies both descriptions within a single framework. This is particularly important for magnetic phenomena, as the unique  structure of Fermi momenta of quarkyonic matter---where quarks occupy the deep Fermi sea while nucleons exist only near the Fermi surface---leads to fundamentally different magnetic responses compared to conventional nuclear matter. The framework enables investigation of spin-dependent interactions across the entire density range relevant to NS cores.

In this work, we develop a framework for studying spin-polarized quarkyonic matter and investigate its magnetic susceptibility. We extend the quarkyonic model to include spin polarization effects and examine how magnetic fields modify the EOS. Our analysis reveals the emergence of ferromagnetic instabilities in pure neutron matter within specific density ranges, with important implications for NS structure and the broader understanding of QCD matter under extreme conditions.


\section{Quarkyonic matter}\label{sec-1}
\begin{figure}[htp]
\centering
\includegraphics[width=0.9\hsize]{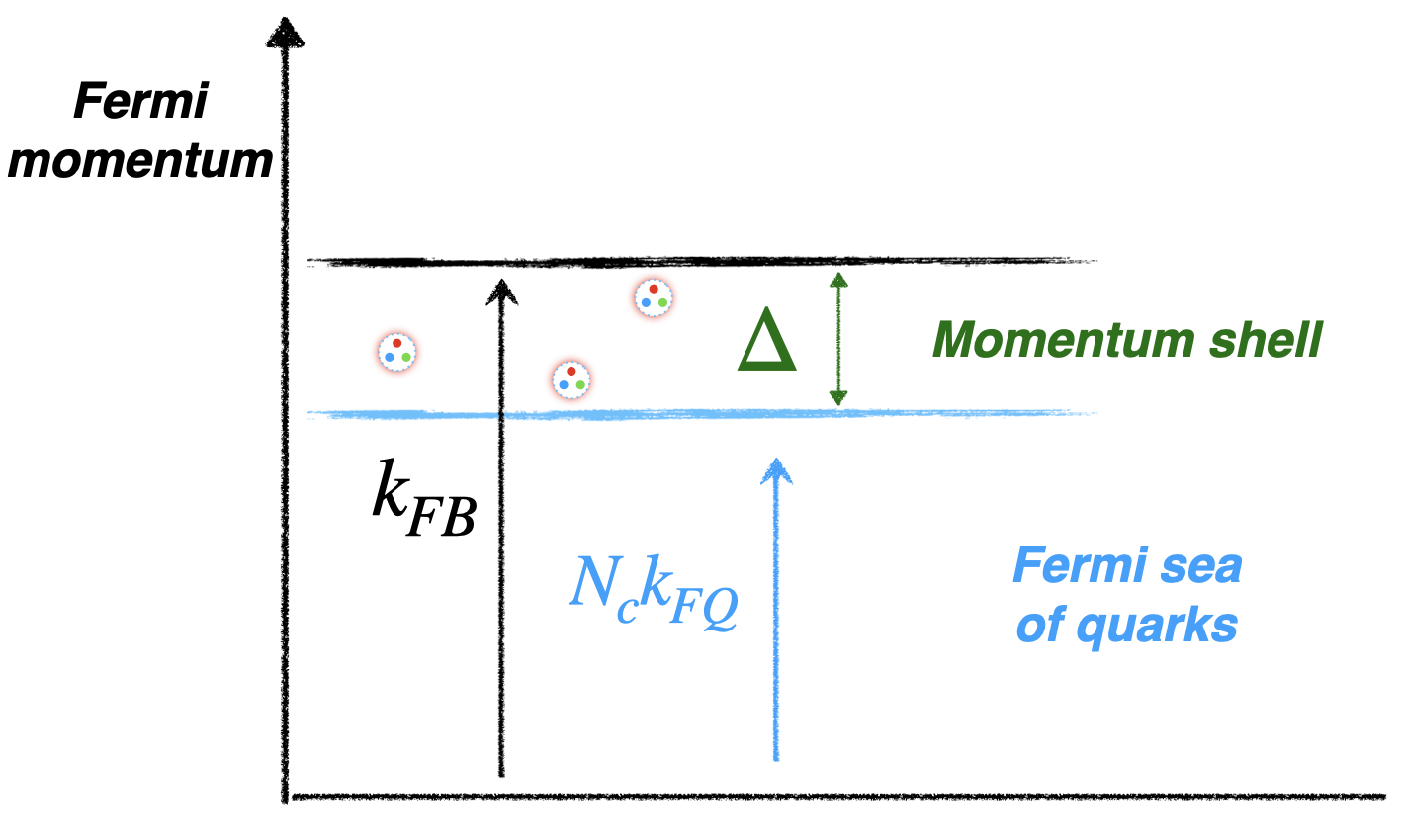}
\caption{Fermi sea structure in momentum space. Quarks occupy the lower momentum states from 0 to $N_c k_{\rm FQ}$, while nucleons exist only in the momentum shell of width $\Delta$ near the Fermi surface.}
\label{fig:Fermi_no_polar}
\end{figure}

Quarkyonic matter represents a novel phase of QCD matter that naturally bridges the gap between nuclear and quark matter at intermediate densities relevant to NS cores. Within this framework, we begin by considering the symmetric matter case and both the nucleon mass and constituent quark mass are held constant vacuum values as $M_N = 939$ MeV and $M_Q = M_N / 3$, respectively. The baryon number density $n_B$ in the model is  expressed as
\begin{align}
n_B = \frac{2}{3\pi^2} \left[ k_{{\rm FB}}^3 - (k_{{\rm FB}} - \Delta)^3 + k_{{\rm FQ}}^3 \right],
\end{align}
where $k_{{\rm FB}}$ is the Fermi momentum of nucleons and the Fermi momentum of quarks $k_{{\rm FQ}}$ is defined as
\begin{align}
k_{{\rm FQ}} = \frac{k_{{\rm FB}}}{N_c}\Theta (k_{{\rm FB}} - \Delta).
\end{align}
with the momentum shell $\Delta$ is parameterized as
\begin{align}
\Delta =  \Lambda_{{\rm QCD}} \left( \frac{\Lambda_{{\rm QCD}}}{ k_{{\rm FB}}} \right)^{\alpha}.
\end{align}
Here the $\Lambda_{{\rm QCD}}$ is the QCD scale parameter, which we set $\Lambda_{{\rm QCD}} = 300$ MeV throughout our study. 
The dimensionless parameter $\alpha$ controls how rapidly the shell width decreases with increasing density. 
This parametrization incorporates previous studies as special cases. For example, $\alpha = 1$ recovers earlier work in~\cite{Zhao:2020dvu} and $\alpha = 2$ corresponds to the form used in~\cite{McLerran:2018hbz}. The total energy is given as
\begin{align}
\varepsilon_N  &= g \int^{k_{{\rm FB}}}_{N_c k_{{\rm FQ}}} \frac{d^3 k}{(2\pi)^3}\sqrt{ k^2 + M_N^2}, \label{eq_nucleon_energy}\\
\varepsilon_Q  &= g N_c \int_0^{k_{{\rm FQ}}} \frac{d^3 k}{(2\pi)^3}\sqrt{ k^2 + M_Q^2}, \label{eq_quark_energy}\\
\varepsilon_{{\rm total}} &= \varepsilon_N + \varepsilon_Q.
\end{align}
Here $\varepsilon_N$ and $\varepsilon_Q$ correspond to the energy densities from nucleons and constituent quarks, respectively. For symmetric nuclear matter, we adopt the spin-isospin degeneracy factor $g=4$. The chemical potential $\mu_B$ and pressure $P$ are then obtained from the thermodynamic relations
\begin{align}
\mu_B &= \frac{\partial \varepsilon_{\rm total}}{\partial n_B},\\
 P& = -\varepsilon_{\rm total} + \mu_B n_B.
\end{align}
At this stage, we do not include any interactions or potentials, treating all nucleons and quarks as free particles (though quarks still obey the Pauli exclusion principle). The corresponding Fermi sea structure is shown in Fig.~\ref{fig:Fermi_no_polar}. In this configuration, quarks occupy the lower momentum states from 0 to $N_c k_{\rm FQ}$, while nucleons exist only near the Fermi surface within the momentum shell $\Delta$.

\begin{figure*}\centering
\includegraphics[width=0.9\hsize]{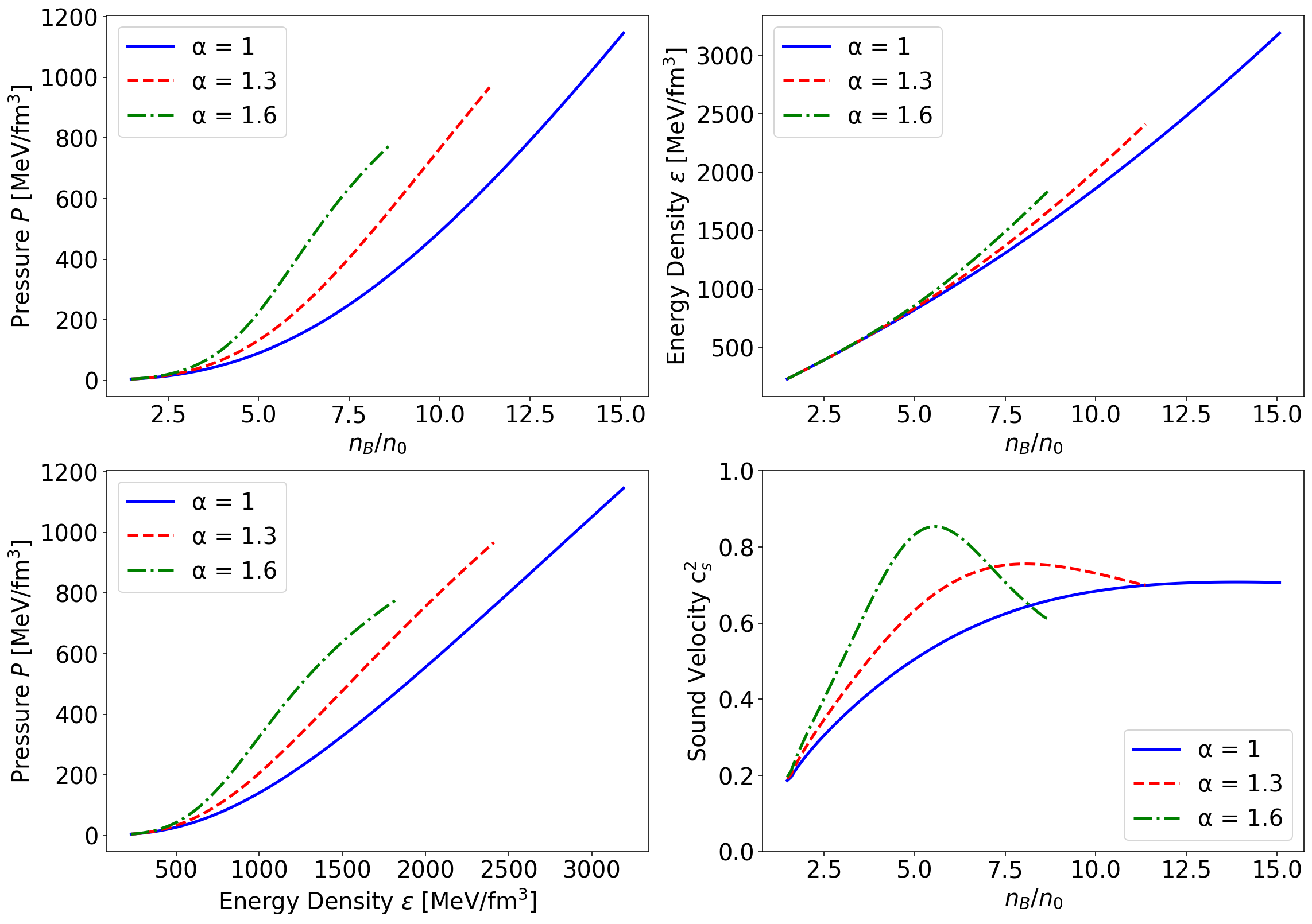}
\caption{Pressure $P$ (upper left) and total energy density $\varepsilon_{\rm total}$ (upper right) as functions of baryon number density; equation of state (lower left) and corresponding sound velocity $c_s^2$ (lower right) for different values of the parameter $\alpha$. }
\label{fig:no_polar}
\end{figure*}

We show the numerical results of thermodynamic quantities in Fig.~\ref{fig:no_polar} for several values of $\alpha$. The pressure $P$ and total energy $\varepsilon_{\rm total}$ as functions of baryon number density are shown in the upper left and upper right panels, respectively. The EOS and the corresponding sound velocity $c_s^2$ are displayed in the lower left and lower right panels.
As $\alpha$ increases, the momentum shell decreases more rapidly, leading to an earlier onset of quark degrees of freedom. Since the pressure is roughly proportional to the number of particles, the appearance of the quark Fermi sea increase the pressure since quark degrees of freedom carry an additional $N_c$ factor, leading to a stiffening of the EOS and a more pronounced sound velocity peak. To be noted, up to this point, we have neglected nuclear interactions and considered only free particle symmetric matter.

\section{Spin polarized quarkyonic matter}

In the presence of a strong magnetic field, particle spins become polarized. We consider symmetric matter and assume that, even after polarization, protons and neutrons retain the same Fermi momentum: The spin-up Fermi momentum is given by $k_{\rm FB}^{\uparrow} = k_p^{\uparrow} = k_n^{\uparrow}$, and the spin-down Fermi momentum by $k_{\rm FB}^{\downarrow} = k_p^{\downarrow} = k_n^{\downarrow}$, where the indices $p$ and $n$ denote protons and neutrons, respectively. 

\begin{figure}[htp]
\includegraphics[width=0.9\hsize]{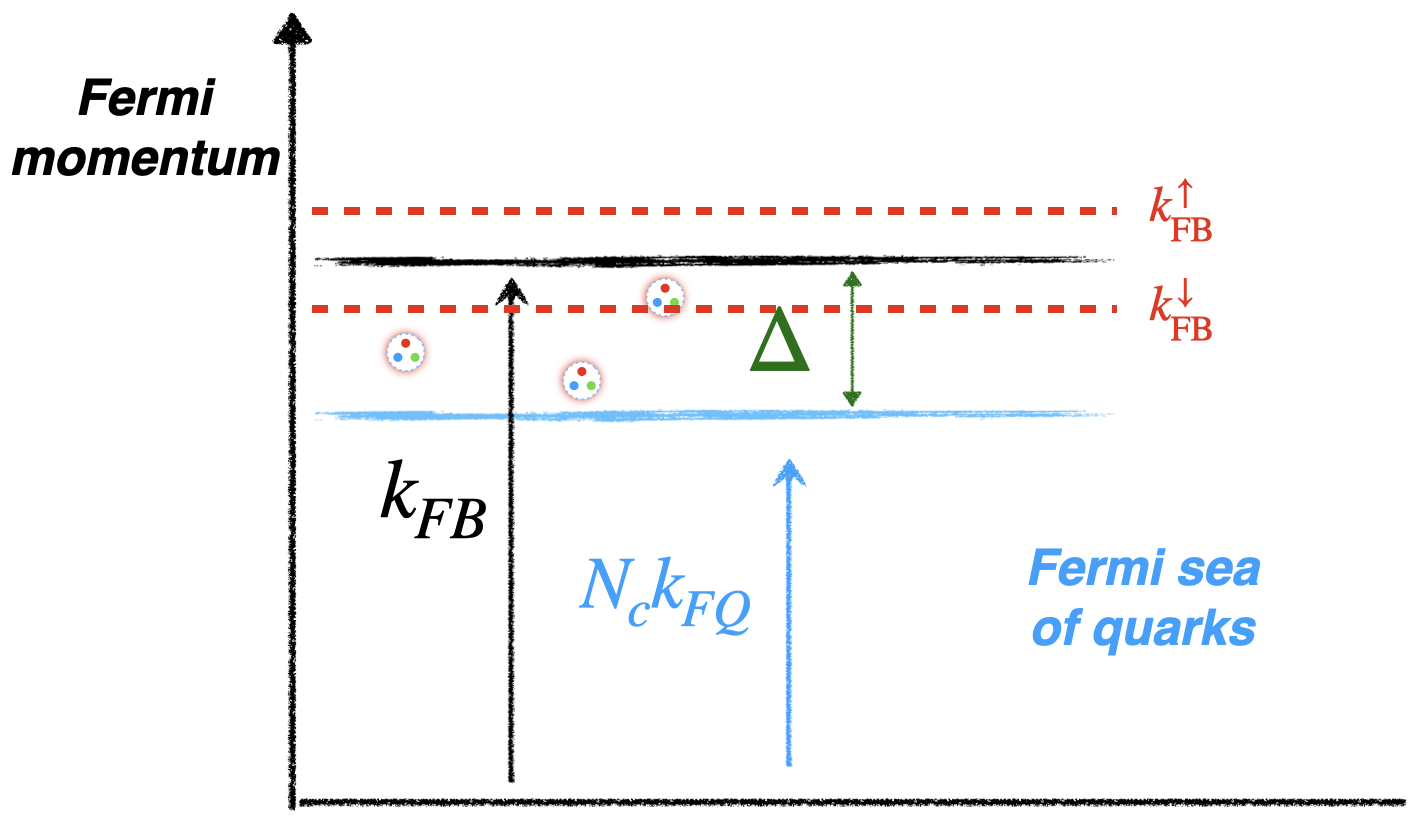}
\caption{Fermi sea structure in momentum space for spin-polarized quarkyonic matter. Quarks occupy the lower momentum states from 0 to $N_c k_{\rm FQ}$, while spin-up and spin-down nucleons exist in separate momentum shells of width $\Delta$ with different Fermi surfaces at $k_{\rm FB}^{\uparrow}$ and $k_{\rm FB}^{\downarrow}$, respectively.  }
\label{fig:Fermi_polar}
\end{figure}

In our treatment, only nucleons are polarized while quarks remain unaffected. This assumption is physically motivated by the distinct characteristics of the two components. The deep quark Fermi sea, composed of weakly interacting quarks at high densities, is highly degenerate and strongly Pauli-blocked. Consequently, it offers limited phase space for spin rearrangement, and spin-dependent interactions are energetically suppressed. In contrast, nucleons in the shell near the Fermi surface can respond to residual spin-dependent nuclear forces.  The corresponding Fermi sea structure is thus modified as shown in Fig.~\ref{fig:Fermi_polar}. Quarks still occupy the lower momentum states from 0 to $N_c k_{\rm FQ}$. For the nucleon part, spin-up nucleons occupy momentum states from $N_c k_{\rm FQ}$ to $k_{\rm FB}^{\uparrow}$, while spin-down nucleons occupy momentum states from $N_c k_{\rm FQ}$ to $k_{\rm FB}^{\downarrow}$.

We can define the dimensionless spin polarization parameter $\xi$ as
\begin{align}
\xi = \frac{ k^{\uparrow}_{\rm FB} -  k^{\downarrow}_{\rm FB}}{2k_{\rm FB}}.
\end{align}
together with the polarized Fermi momenta $k^{\uparrow}_{\rm FB}$ and $k^{\downarrow}_{\rm FB}$ as
\begin{align}
k^{\uparrow}_{\rm FB} = (1+\xi)k_{\rm FB}, \quad k^{\downarrow}_{\rm FB} = (1 - \xi) k_{\rm FB}.
\end{align}

The energy density for the nucleonic component is then modified to
\begin{align}
\varepsilon_N =& 2\int_{N_c k_{\rm FQ}}^{ k^{\downarrow}_{\rm FB} } \frac{d^3 k}{ (2\pi)^3}\sqrt{k^2 + M_N^2} \\
&+ 2\int_{N_c k_{\rm FQ}}^{ k^{\uparrow}_{\rm FB} } \frac{d^3 k}{ (2\pi)^3}\sqrt{k^2 + M_N^2}.
\end{align}
Here we consider only the isospin degeneracy, resulting in a factor of 2 for each integration. When $\xi = 0$, the energy density coincides with the previous form in Eq.~(\ref{eq_nucleon_energy}). The energy density from the quark component still takes the same form as in Eq.~(\ref{eq_quark_energy}). The baryon number density now becomes
\begin{align}
n_B =& \frac{2}{3\pi^2}k_{\rm FQ}^3 +  \frac{1}{3\pi^2} \left[ (k^{\uparrow}_{\rm FB})^3 - (k^{\uparrow}_{\rm FB} - \Delta)^3 \right.\nonumber\\
&\left.   +  (k^{\downarrow}_{\rm FB})^3 - (k^{\downarrow}_{\rm FB} - \Delta)^3  \right]
\end{align}

\begin{figure*}\centering
\includegraphics[width=0.9\hsize]{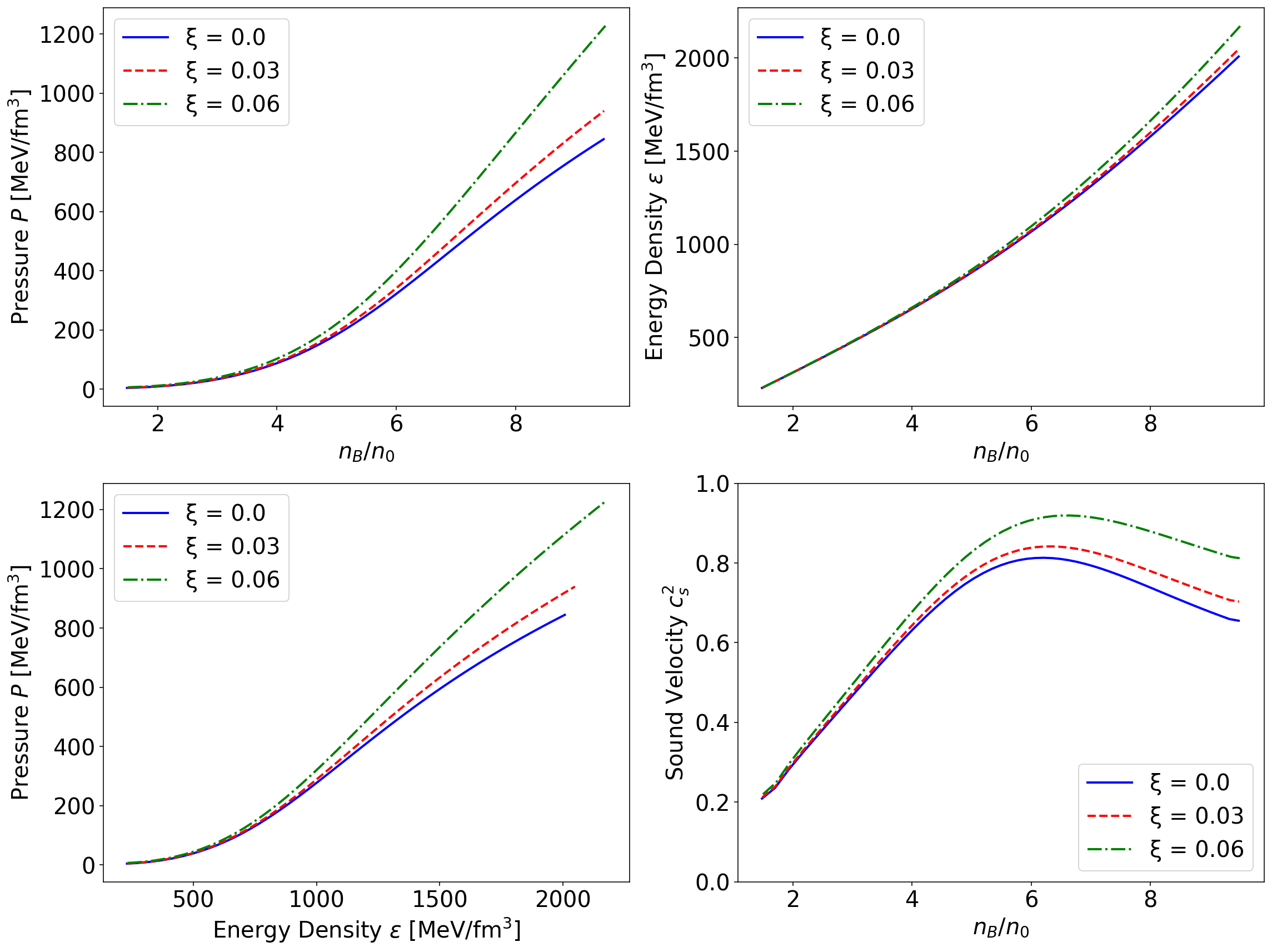}
\caption{Pressure $P$ (upper left) and total energy density $\varepsilon_{\rm total}$ (upper right) as functions of baryon number density; equation of state (lower left) and corresponding sound velocity $c_s^2$ (lower right) for spin-polarized quarkyonic matter. The momentum shell parameter is fixed at $\alpha = 1.5$, while the spin polarization parameter varies as $\xi = 0, 0.03, 0.06$. }
\label{fig:polar}
\end{figure*}

We then calculate the baryon chemical potential $\mu_B$ and pressure $P$ following the previous section. The results are shown in Fig.~\ref{fig:polar}, where we fix $\alpha = 1.5$ and vary the polarization parameter as $\xi = 0, 0.03, 0.06$. 

Upon spin polarization, both the pressure and energy increase at the same baryon density. This behavior arises because spin polarization breaks the degeneracy between spin-up and spin-down states, effectively reducing the available phase space and increasing the kinetic energy of the system. Consequently, the EOS becomes stiffer and the sound velocity increases. The magnitude of these effects scales with the polarization parameter $\xi$, which in realistic NS conditions would be determined by the balance between magnetic field strength and thermal effects. The values of $\xi$ considered here ($0.03$ to $0.06$) correspond to moderate spin polarization levels that could be realized in strongly magnetized neutron star environments (magnetars), depending on the microscopic interactions of dense matter. \cite{Maruyama:2000cw,Bordbar:2011rt,Dong:2013hta}. 

To describe NS matter, we need to impose the constraints of local charge neutrality and beta equilibrium. Charge neutrality requires that the total electric charge vanishes, while beta equilibrium ensures that weak interaction processes (such as neutron $\beta$-decay and electron capture) are in balance. These conditions restrict the proton fraction to less than 10\%. For this reason, we approximate the matter as consisting entirely of neutrons. At a given baryon density $n_B$, we define the unpolarized neutron Fermi momentum as $k_{\rm FB}$, while the polarized neutron Fermi momenta are denoted by $k_{FB}^{\uparrow}$ and $k_{FB}^{\downarrow}$ for spin-up and spin-down neutrons, respectively. Similarly, the up and down quark Fermi momenta are represented by $k_{\rm Fu}$ and $k_{\rm Fd}$. The charge neutrality constraint establishes the following relationships between the quark Fermi momenta:
\begin{align}
k_{\rm Fd} = \frac{k_{\rm FB} - \Delta}{3}, \quad k_{\rm Fu} = \frac{k_{\rm Fd}}{2^{1/3}}.
\end{align}
These relations ensure that the electric charges of up quarks (+2/3) and down quarks (-1/3) balance appropriately.
Realistic calculations of neutron star equations of state have demonstrated the importance of neutron-neutron interactions. The nature of these interactions depends strongly on density: at low densities where $n_n \leq n_0$ (where $n_0 \approx 0.16$ fm$^{-3}$ is nuclear saturation density), interactions are predominantly attractive and serve to reduce the pressure of the neutron matter. This attractive regime is dominated by the residual strong force between neutrons. However, as density increases beyond nuclear saturation, repulsive two- and three-body interactions at short distances become dominant, leading to a rapid pressure increase. This repulsive core is essential for supporting neutron stars against gravitational collapse.

To incorporate these interaction effects, we adopt the parametrization from Ref.~\cite{Gandolfi:2013baa}, which provides a simple but accurate fit to sophisticated microscopic neutron matter calculations based on realistic nuclear forces. For densities $n_n \leq 2n_0$, the interaction energy density is well approximated by

\begin{align}
V_n (n_n) =& \tilde{a} n_n \left(\frac{n_n}{n_0} \right) + \tilde{b} n_n \left( \frac{n_n}{n_0} \right)^2 \\
 &+ \tilde{p} (n_n^{\uparrow} - n_n^{\downarrow})^2. \label{p_term}
\end{align}
Here the coefficients $\tilde{a} = -28.6 \pm 1.2$ MeV and $\tilde{b} = 9.9 \pm 3.7$ MeV are chosen to bracket the uncertainties due to poorly constrained three-neutron forces. The negative value of $\tilde{a}$ reflects the attractive nature of neutron interactions at low densities, while the positive $\tilde{b}$ coefficient captures the repulsive behavior at higher densities.

The additional $\tilde{p}$ term in Eq.~(\ref{p_term}) represents the energy contribution from spin polarization, proportional to the square of the neutron spin asymmetry $(n_n^{\uparrow} - n_n^{\downarrow})^2$. This term arises from spin-dependent interactions between neutrons, including spin-orbit coupling, tensor forces, and exchange interactions. The sign of $\tilde{p}$ determines whether polarization is energetically favorable (negative $\tilde{p}$) or costly (positive $\tilde{p}$). Note that $\tilde{p}$ is less\sout{well}-constrained experimentally, as it requires knowledge of spin-dependent nuclear interactions at high densities. With the interaction potential defined, we obtain the number densities for neutrons and quarks as
\begin{align}
n_n &= \frac{1}{6\pi^2}  \left[ (k^{\uparrow}_{\rm FB})^3 - (k^{\uparrow}_{\rm FB} - \Delta)^3 \right.  \nonumber \\
& \left.\quad +  (k^{\downarrow}_{\rm FB})^3 - (k^{\downarrow}_{\rm FB} - \Delta)^3  \right], \\
n_Q &= \frac{ k_{{\rm Fd}}^3 + k_{{\rm Fu}}^3 }{3\pi^2}.
\end{align}
The total baryon number density is $n_B = n_n + n_Q$. The energy density of the polarized quarkyonic matter is 
\begin{align}
\varepsilon_N =& \int_{N_c k_{\rm FQ}}^{ k^{\downarrow}_{\rm FB} } \frac{d^3 k}{ (2\pi)^3}\sqrt{k^2 + M_N^2} \nonumber \\
&+ \int_{N_c k_{\rm FQ}}^{ k^{\uparrow}_{\rm FB} } \frac{d^3 k}{ (2\pi)^3}\sqrt{k^2 + M_N^2} + V_n (n_n), \\
\varepsilon_Q =&2\sum_{i = u, d} N_c \int_0^{k_{{\rm Fi}}} \frac{d^3 k}{(2\pi)^3}\sqrt{k^2 + M_Q^2}, \\
\varepsilon_{{\rm total}} =&\varepsilon_N + \varepsilon_Q.
\end{align}

The chemical potential and pressure follow from standard thermodynamic relations: $\mu_B = \partial \varepsilon_{{\rm total}} / \partial n_B$ and $P = -\varepsilon_{{\rm total}} + \mu_B n_B$. To characterize the magnetic response of the system, we define the spin susceptibility as
\begin{align}
\chi = \frac{\partial^2 \varepsilon_{{\rm total}}}{\partial \xi^2} \bigg|_{\xi=0}.
\end{align}
This quantity measures the system's resistance to spin polarization: a positive $\chi$ indicates that external magnetic fields are required to induce polarization, with larger values corresponding to greater resistance. The spin susceptibility directly determines the critical field strength needed for significant polarization in neutron star environments and provides insight into potential ferromagnetic instabilities (negative $\chi$) that could drive spontaneous phase transitions in dense matter. 
\begin{figure*}\centering
\includegraphics[width=0.9\hsize]{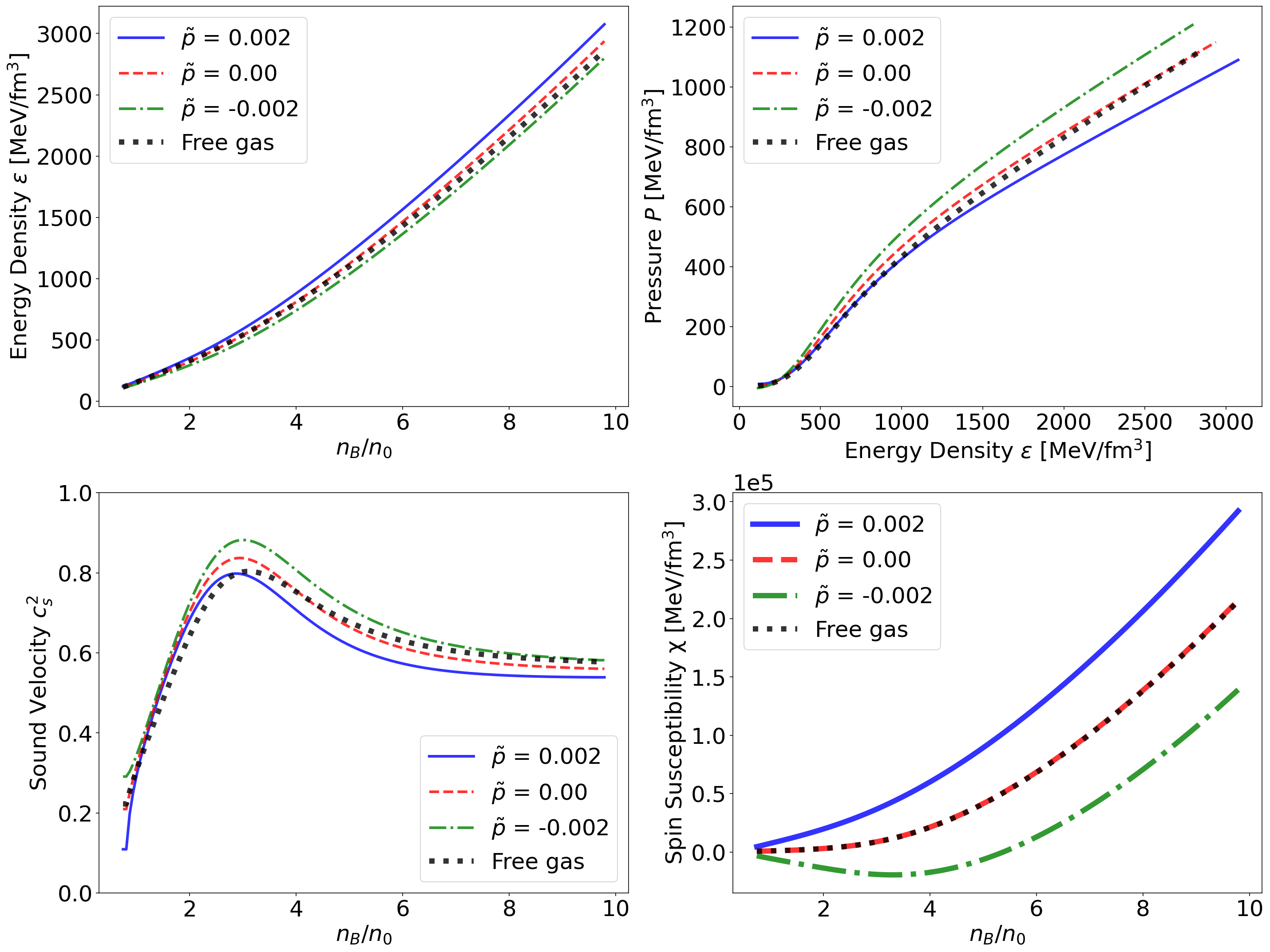}
\caption{Total energy density $\varepsilon_{\text{total}}$ (upper left) and equation of state (upper right) as functions of baryon number density; sound velocity $c_s^2$ (lower left) and spin susceptibility $\chi$ (lower right) for spin-polarized quarkyonic matter. The momentum shell parameter is fixed at $\alpha = 1.5$ and the spin polarization parameter is fixed as $\xi = 0.06$, while the spin-dependent interaction parameter varies as $\tilde{p} = 0.002$, $0$, and $-0.002$ MeV$\cdot$fm$^6$.}
\label{fig:p_term}
\end{figure*}

We present the numerical results for polarized pure neutron matter in Fig.~\ref{fig:p_term}, where we examine the effects of the spin-dependent interaction parameter $\tilde{p}$. The calculations are performed with fixed parameters $\alpha=1.5$ and $\xi=0.06$, while varying $\tilde{p}$ across different values: $\tilde{p} = 0.002$ MeV$\cdot$fm$^6$ (green curve), $0$ (red curve), and $-0.002$ MeV$\cdot$fm$^6$ (blue curve). We also show the results for the free Fermi gas of neutrons (blck dotted curve) for comparison. The figure displays the energy density versus baryon number density (upper left panel), the equation of state (upper right panel), sound velocity as a function of baryon number density (lower left panel), and spin susceptibility $\chi$ versus baryon number density (lower right panel). 

The results demonstrate the expected behavior of the spin-dependent interaction term. Positive values of $\tilde{p}$ increase the total energy density, reflecting an energetic cost for spin polarization, while negative values decrease the energy density, indicating that spin polarization becomes energetically favorable. This behavior directly follows from the quadratic dependence on spin asymmetry in Eq.~(\ref{p_term}). Also as the energy density decrease, the EOS becomes stiffer accompanied by a corresponding increase in the sound velocity. 

Of particular interest is the behavior of the spin susceptibility $\chi$.  When $\tilde{p} = -0.002$ MeV$\cdot$fm$^6$, we observe that $\chi$ becomes negative at densities below approximately $5.5n_0$ before returning to positive values at higher densities. 
At low densities, the negative $\tilde{p}$ term in the  potential provides an attractive contribution proportional to $(n_n^{\uparrow} - n_n^{\downarrow})^2$, which energetically favors spin polarization. When this spin-dependent attraction dominates over the kinetic energy cost of breaking spin degeneracy, the system develops a ferromagnetic instability characterized by negative $\chi$. This indicates that the matter would spontaneously develop spin polarization even without external magnetic fields, similar to ferromagnetic phase transitions in condensed matter systems. The magnitude of this effect depends on the balance between the interaction strength $|\tilde{p}|$ and the kinetic energy density of the system.

As density increases beyond $5.5n_0$, the kinetic energy of the neutron matter grows more rapidly than the interaction energy, eventually overwhelming the attractive spin-dependent force. The Pauli pressure associated with maintaining different Fermi momenta for spin-up and spin-down neutrons becomes prohibitively large, forcing the system back into a paramagnetic state with positive $\chi$. This transition represents a phase transition from ferromagnetic to paramagnetic behavior driven purely by density effects.

A particularly noteworthy finding is that this ferromagnetic instability occurs in pure neutron matter within the quarkyonic framework. Previous investigations\cite{Dong:2013hta}, have primarily attributed negative magnetic susceptibility and potential ferromagnetic behavior to the contributions from proton interactions in $\beta$-equilibrium matter. In their study of magnetization of neutron star matter, they found that protons, rather than electrons, contribute most significantly to the magnetization in relatively weak magnetic fields, and emphasized the role of charged particle interactions in generating strong magnetic responses.

However, our results demonstrate that the quarkyonic matter structure itself can generate ferromagnetic instabilities in pure neutron matter, independent of proton contributions if we have 
an attractive spin-spin interaction between neutrons. When we extend our calculations to include $\beta$-equilibrium with protons and leptons, the fundamental tendency toward negative $\chi$ at intermediate densities should not be changed, since the underlying physics driving this instability remains dominated by the neutron component. In $\beta$-equilibrium neutron star matter, the proton fraction is constrained by charge neutrality and weak equilibrium conditions to remain below approximately 10\% throughout the density range of interest. Consequently, the quarkyonic momentum shell structure continues to be primarily populated by neutrons, and the spin-dependent interaction term $\tilde{p}(n_n^{\uparrow} - n_n^{\downarrow})^2$ that drives the ferromagnetic instability remains the dominant contribution to the magnetic susceptibility.

\section{Summary and Discussion}\label{sec-summary}

In this work, we have developed a framework for studying magnetized quarkyonic matter and investigated its magnetic susceptibility properties. We extended the quarkyonic matter model to include spin polarization effects, treating nucleons near the Fermi surface as spin-polarizable while keeping quarks in the deep Fermi sea unpolarized due to strong Pauli blocking. 

Our calculations reveal the emergence of ferromagnetic instabilities in pure neutron matter within specific density ranges. For negative values of the spin-dependent interaction parameter $\tilde{p}=-0.02$ MeV$\cdot$fm$^6$, the magnetic susceptibility $\chi$ becomes negative at densities below approximately $5.5n_0$, indicating spontaneous spin polarization even without external magnetic fields. As density increases further, the system transitions back to paramagnetic behavior when kinetic energy costs overwhelm the attractive spin-dependent interactions. 

A particularly noteworthy finding is that this ferromagnetic instability occurs in pure neutron matter within the quarkyonic framework, independent of proton contributions. Previous investigations have primarily attributed negative magnetic susceptibility and ferromagnetic behavior to proton interactions in $\beta$-equilibrium matter. Our results demonstrate that the quarkyonic matter structure itself can generate ferromagnetic instabilities, suggesting a new mechanism for magnetism in dense matter.

When extended to $\beta$-equilibrium NS matter, the fundamental tendency toward negative $\chi$ at intermediate densities should persist, since the underlying physics driving the ferromagnetic instability remains dominated by the neutron component. The proton fraction in $\beta$-equilibrium matter is constrained to below approximately 10\% throughout the density range of interest, ensuring that the quarkyonic momentum shell structure continues to be primarily populated by neutrons.

Several important extensions of this work merit future consideration. A full treatment of $\beta$-equilibrium matter including protons, electrons, and muons would provide more realistic neutron star conditions. The incorporation of temperature effects could reveal the critical conditions for ferromagnetic-paramagnetic transitions relevant to neutron star cooling and thermal evolution. The spin-dependent interaction parameter $\tilde{p}$ remains poorly constrained experimentally, highlighting the need for better theoretical and experimental determinations of spin-dependent nuclear interactions at high densities. Additionally, the connection between our microscopic magnetic susceptibility calculations and macroscopic magnetic field evolution in NSs requires further investigation through magnetohydrodynamic modeling.

In conclusion, we have demonstrated that quarkyonic matter exhibits rich magnetic phenomenology, including ferromagnetic instabilities that arise from the unique momentum-space structure inherent to this phase of dense matter. The ability to generate ferromagnetism in pure neutron matter represents a significant departure from conventional nuclear matter physics and opens new avenues for understanding NS magnetism. Our results constitute an important step toward understanding the magnetic properties of matter under extreme conditions and provide a theoretical foundation for interpreting the diverse magnetic phenomena observed in neutron stars and magnetars.


\section*{acknowledgments}
This work was supported by the JSPS KAKENHI (Grant Nos. JP23H05434, JP25H01269, JP25H00402 and JP25K07322), 
and JST ERATO (Grant No. JPMJER2304).


\bibliography{ref.bib}

\end{document}